\documentclass[a4paper,twocolumn]{article}
\usepackage{multicol}
\usepackage{mathrsfs}
\usepackage{amsmath}
\usepackage{graphicx}
\usepackage[left=1in,right=1in,top=.8in,bottom=.8in]{geometry}
\usepackage{hyperref}
\usepackage[noblocks]{authblk}

\vspace{10cm}
\date{}
\begin{document}

\title{\textbf{Thermodynamic Properties of Ice: A Monte Carlo Study}}

\author{Tanmay Bhore \thanks{}}
\author{Siddharth Chaini\thanks{}}
\author{Siddharth Bachoti\thanks{}}
\author{Vipin Khade\thanks{}}
\author{Vinay Patil\thanks{}}
\affil{Department of Physics, Indian Institute of Science Education and Research Bhopal, Bhopal, 462066}


\twocolumn[
	\begin{@twocolumnfalse}
		\maketitle
	\begin{abstract}
	In this text, we implement a monte carlo algorithm to study thermodynamic properties of ice. Our program, written in Python, is open-sourced and available on \href{https://github.com/AKnightWing/ColdAsIce}{Github}\footnotemark. We develop a novel scheme to compute the residual entropy of a two dimensional lattice model of ice at absolute zero. A model of energetic ice is also considered and its thermodynamic properties are studied with temperature. We report a remarkable result for the energetic ice model, the presence of a phase transition at non zero temperatures.
	\end{abstract}
	\end{@twocolumnfalse}
	]
{
	\renewcommand{\thefootnote}%
	{\fnsymbol{footnote}}
	\footnotetext[1]{tanmayb17@iiserb.ac.in}
	\footnotetext[2]{siddharthc17@iiserb.ac.in}
	\footnotetext[3]{siddharthb17@iiserb.ac.in}
	\footnotetext[4]{vipink17@iiserb.ac.in}	
	\footnotetext[5]{vinay17@iiserb.ac.in}
}

	
\section{Introduction}
\footnotetext{\href{https://github.com/AKnightWing/ColdAsIce}{https://github.com/AKnightWing/ColdAsIce}}

$H_2O$ in its solid form, is found abundantly in nature. Therefore it is only natural that its structure has been of computational interest over the last century.
\\ \\In a water molecule, each oxygen atom shares a covalent bond with two hydrogen atoms and forms hydrogen bonds with two more hydrogen atoms. The protons are located between pairs of nearest neighbour oxygen atoms. The potential energy of such a proton has two possible minima as a function of distance on the line joining two oxygen atoms. In hexagonal ice, two oxygen atoms are 2.75{\AA} apart, and the most favourable position for each hydrogen atom is to be at 1{\AA} from each oxygen atom. Thus each proton has to choose between two positions 0.75{\AA} apart. 
To build a coherent model of ice, we must make sure that the arrangement described above holds.

\section{Ice Rules}
To satisfy the conditions given in Section 1.; Bernar, Fowler and Pauling proposed the following rules for the arrangement of protons on bonds of the lattice:
\begin{enumerate}
	\item There is precisely one hydrogen atom on each hydrogen bond.
	\item There are precisely two hydrogen atoms near each oxygen atom.
\end{enumerate}
These two constraints are collectively known as the \textbf{ice rules}. In real ice, at finite temperatures, ice rules do not necessarily hold. For example, the first ice rule is violated if a hydrogen bond is populated by 0 or 2 protons. These defects are energetically highly unfavourable and only have a significant contribution to the overall structure at higher temperatures. 
\\ The second ice rule is violated if one or three protons neighbour each oxygen. However, this defect is even more energetically unfavourable than the first and thus occurs rarely at lower temperatures.
\\Keeping in mind the relative frequency of occurence of these defects, we study a class of models called \textbf{ice models} which follow the ice rules exactly. At lower temperatures, the two ice rules serve as accurate constraints that provide a framework for calculating thermodynamic properties of ice.

\section{Residual Entropy}
In a micro-canonical ensemble, the entropy of the system is defined as $S = k_B \ln(\Omega)$ where $\Omega$ is the number of microstates of the system. Many systems studied in physics have a unique ground state - the state in which energy of the system is minimum. 
\\Sometimes a system may have several degenerate ground states, such as the Ising model, in which case the system tends to have a non zero entropy at absolute zero since entropy goes as the logarithm of number of degenerate states.
\\Ice, is special, in the sense that it has many possible configurations dictated by the ice rules, and all these configurations have roughly the same energy. So, there are as many ground states as there are possible configurations of protons. This number increases exponentially with system size, which means that even in the thermodynamic limit, the system tends to have a non zero entropy at zero temperature, known as \textbf{Residual Entropy}.

\section{Monte Carlo Simulation}
Consider a dynamic physical system in the state $\mu$. If we consider some quantity $Q$ which takes the value $Q_{\mu}$ for the system in state $\mu$, then we define the expectation value of $Q$ at a given time $t$ as
\begin{equation}
\langle Q \rangle = \sum_{\mu} Q_{\mu} w_{\mu}(t)
\end{equation}
Now for the square ice model, the number of possible states of a system increases rapidly with lattice size and thus it's not computationally favourable to sample all possible states of a system. Monte Carlo techniques work by choosing a subset of states at random from some probability distribution $p_{\mu}$ which we specify. Suppose we choose to sample through $M$ such states, then our best estimate of a quantity we're trying to measure is:
\begin{equation}\label{sampling}
Q_M = \frac{\sum_{i = 1}^{M }Q_{\mu_{i}}p_{\mu_i}^{-1}e^{-\beta E_{\mu_i}}}{\sum_{j=1}^{M} p_{\mu_{j}}^{-1}e^{-\beta E_{\mu_j}}}
\end{equation}
where $\beta = 1/k_B T$
\subsection{Importance Sampling}
The question is, what probability distribution do we choose? We know that physical systems sample possible states with the Boltzmann probability distribution. With this in hindsight, we choose the probability that a particular state $\mu$ gets chosen for sampling as $p_{\mu} = Z^{-1}e^{-\beta E_{\mu}}$, where $Z$ is the partition function given by:
\begin{equation}
Z = \sum_{\mu}e^{-\beta E_{\mu}}
\end{equation}
Then, Eq.(\ref{sampling}) becomes:
\begin{equation}
Q_M =  \frac{1}{M} \sum_{i=1}^{M} Q_{\mu_i}
\end{equation}
\subsection{Markov Processes}
A Markov process is a mechanism  which, given a system in state $\mu$, generates a new state of that system $\nu$. The probability of generating the state $\nu$ from $\mu$ is called the \textbf{Transition Probability}: $P(\mu \rightarrow \nu)$. All transition probabilities in a Markov process must satisfy:\\
\\1. All P's are time independent.
\\2. They should depend only on the properties of the current states $\mu$ and $\nu$ and not on any other states the system has passed through.
This ensures that the probability of generating $\nu$ from $\mu$ is the same no matter where the system has come from. \\ \\
A constraint on the transition probabilities is  $\sum_{\nu}P(\mu \rightarrow \nu) = 1$
\subsection{Ergodicity}
The condition of ergodicity is the requirement that it should be possible for our markov process to reach any state of the system from any other state, if we run the process long enough.\\
Note: The Markov process is choosen specifically so that when we run it for long enough, starting from any state of a system, we will get a succession of states which appear with probabilities given by the Boltzmann distribution.
\subsection{Detailed Balance}
This is a condition we impose on our Markov process to ensure that at equilibrium, the system follows the Boltzmann probability distribution. At equilibrium, the rate at which a system makes transitions in and out of any state must be equal. Thus\\ $\sum_{\nu} p_{\nu}P(\mu \rightarrow \nu) = \sum_{\nu}p_{\nu}P(\nu\rightarrow\mu)$. We also know that $\sum_{\nu}P(\mu \rightarrow \nu) = 1$. Hence 
\begin{equation}
p_{\mu} = \sum_{\nu}p_{\nu} P(\nu \rightarrow \mu)
\end{equation}
For any set of transition probabilities that satisfy Eq.(9), $p_{\mu}$ will be the probability distribution in equilibrium for the Markov process. We state without proof the desired condition for detailed balance in our case\cite{Barkema_1998}:
\begin{equation}
\frac{P(\mu \rightarrow \nu)}{P(\nu \rightarrow \mu)} = \frac{p_{\nu}}{p_{\mu}} = e^{-\beta(E_{\nu}-E_{\mu})}
\end{equation}
\subsection{Acceptance Ratios}
Write $P(\mu \rightarrow \nu) = g(\mu \rightarrow \nu)A(\mu \rightarrow \nu)$ where $g(\mu \rightarrow \nu)$ is the \textit{selection probability}. It is the probability that given an initial state $\mu$, our algorithm generates a new state $\nu$, and $A(\mu \rightarrow \nu)$ is the \textit{Acceptance Ratio}. It gives the probability of accepting a new state generated by our algorithm. Then, $A(\mu \rightarrow \nu) \in [0,1]$
\\Thus, Eq.(10) becomes
\begin{equation}
\frac{P(\mu \rightarrow \nu)}{P(\nu \rightarrow \mu)} = \frac{A(\mu \rightarrow \nu)}{A(\nu \rightarrow \mu)}\frac{g(\mu \rightarrow \nu)}{g(\nu \rightarrow \mu)}
\end{equation}

\section{Data Representation}
\begin{figure}[h]
	\centering
	\includegraphics[scale=0.55]{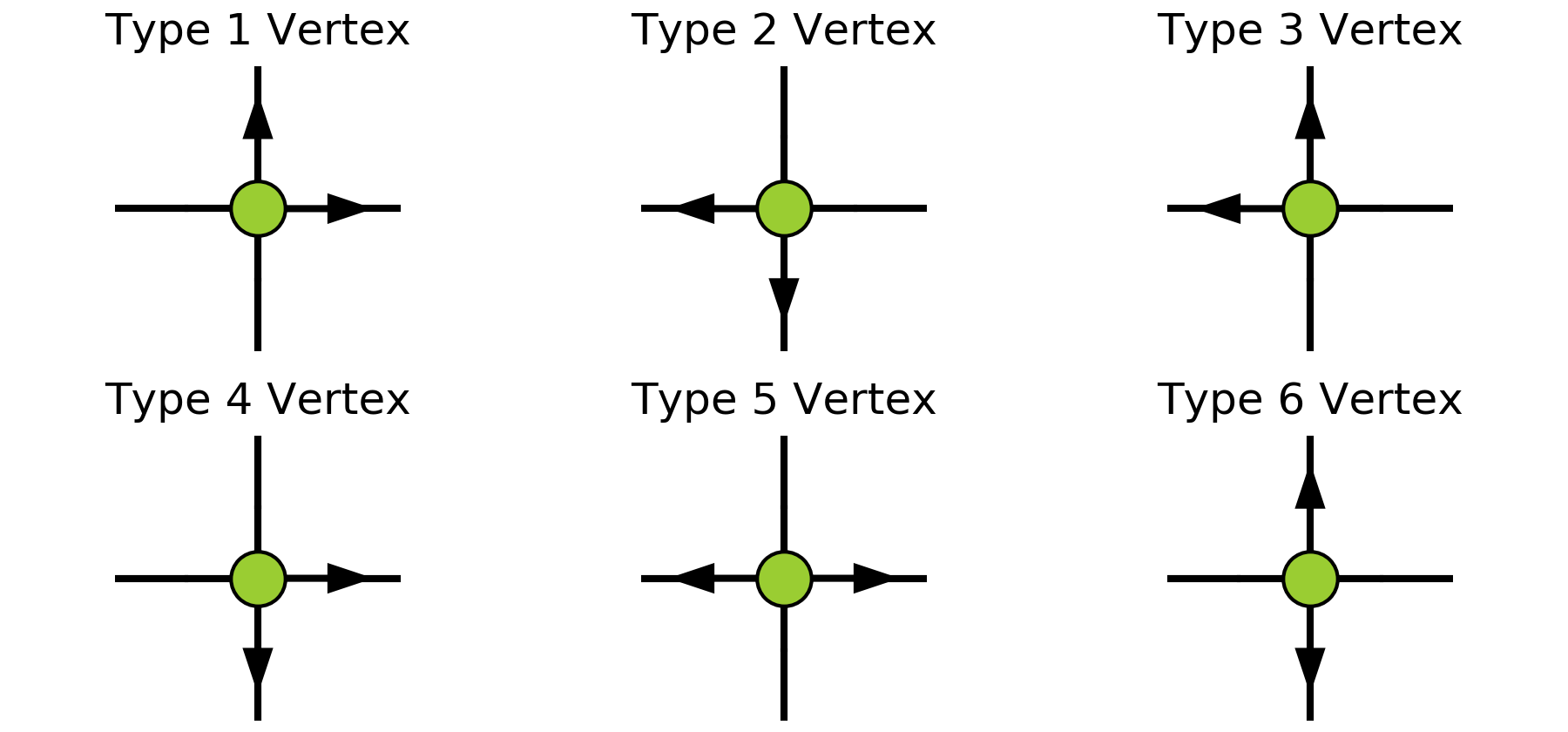}
	\caption{Six Possible vertex configurations}
	\label{fig:6vertices}
\end{figure}

\begin{figure}[h]
	\centering
	\includegraphics[scale=0.132]{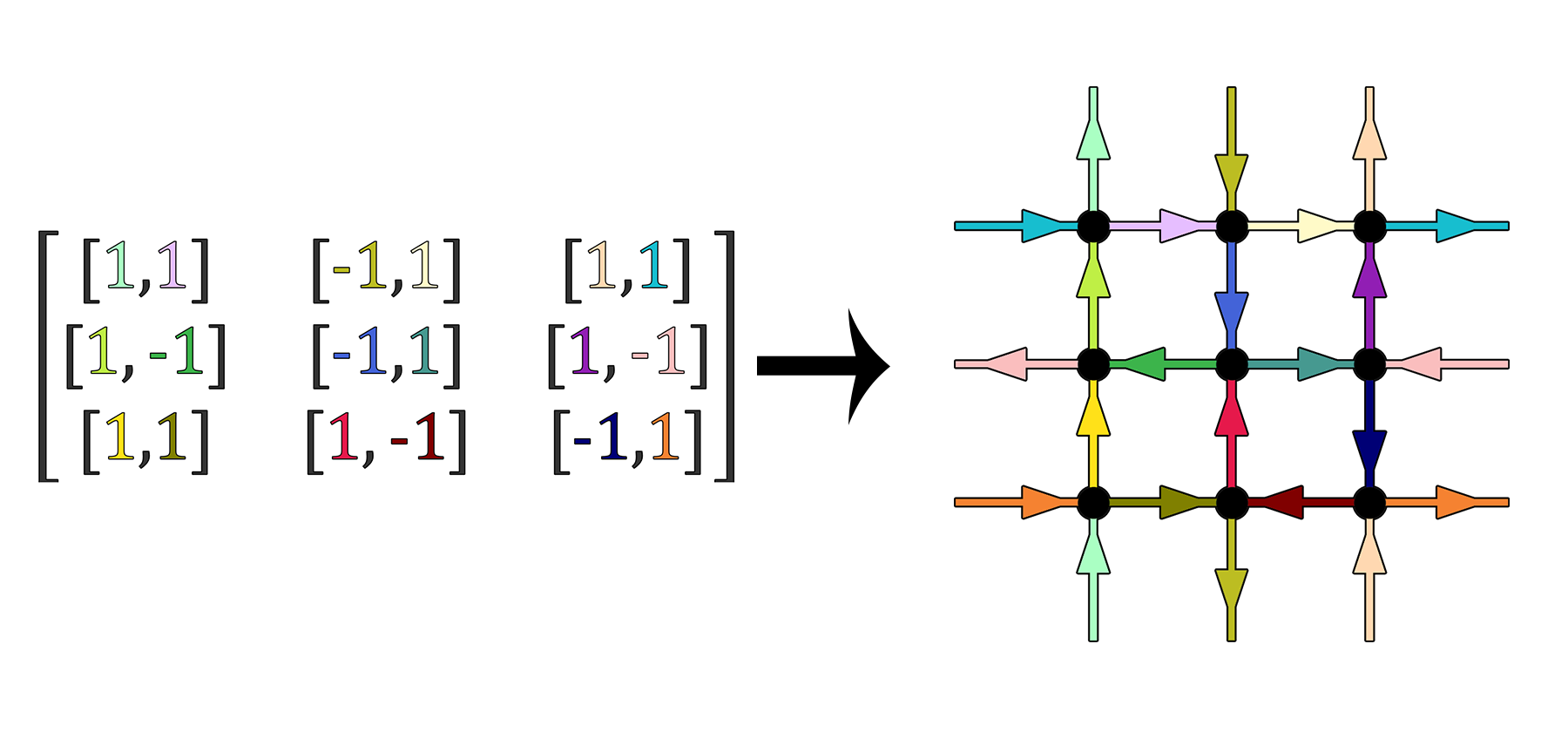}
	\caption{An example of data representation with periodic boundaries}
	\label{fig:datarep}
\end{figure}

This section describes the data representation techniques we have used hereon.
\newline To construct a physical model out of the ice rules, we will consider a lattice of oxygen atoms placed on a 2-D plane. In Square Ice, oxygen atoms are arranged on the vertices of a square lattice. Thus, each oxygen atom has a hydrogen bond on all four sides, which are represented by the lines of the grid. Protons, which reside on bonds, are represented by arrows.
\\The direction of the arrow is such that it points towards the end of the bond nearest to the proton. This arrangement thus clearly satisfies the condition requiring oxygen to have coordination number equal to four. From the second ice rule, we further get the condition that each oxygen atom must have exactly two arrows pointing towards it, and two arrows pointing away from it. We thus conclude that for a singular lattice point, there are 6 possible configurations as given in Fig.(\ref{fig:6vertices}). 
\\Further, we make the assumption that the \textbf{boundaries are periodic}. This means that the cell satisfies perfect two-dimensional tiling. This point will be crucial in the symmetry analysis later on. \textit{Because of the above restrictions, it is sufficient to label each $(i, j)^{th}$ oxygen atom with the configuration of the bond above and right to it}. The details of the bond below and left of the current atom can be determined from the labels of the $(i+1, j)^{th}$ and $(i, j-1)^{th}$ atoms respectively. For these 2 labels, we have used the convention where +1 denotes that the direction of the arrow is pointing away from the atom while -1 denotes that it is pointing towards the atom. Thus, for a grid of n $\times$ n, we have an array of shape (n, n, 2) when using our convention to label it. An example is given in Fig.(\ref{fig:datarep}). \newline This framework is thus capable of uniquely describing any possible configuration for an ice model and also makes visualising the lattice possible through a Python program.

\section{Long loop algorithm}
\begin{figure*}[h]
	\centering
	\includegraphics[scale=0.28]{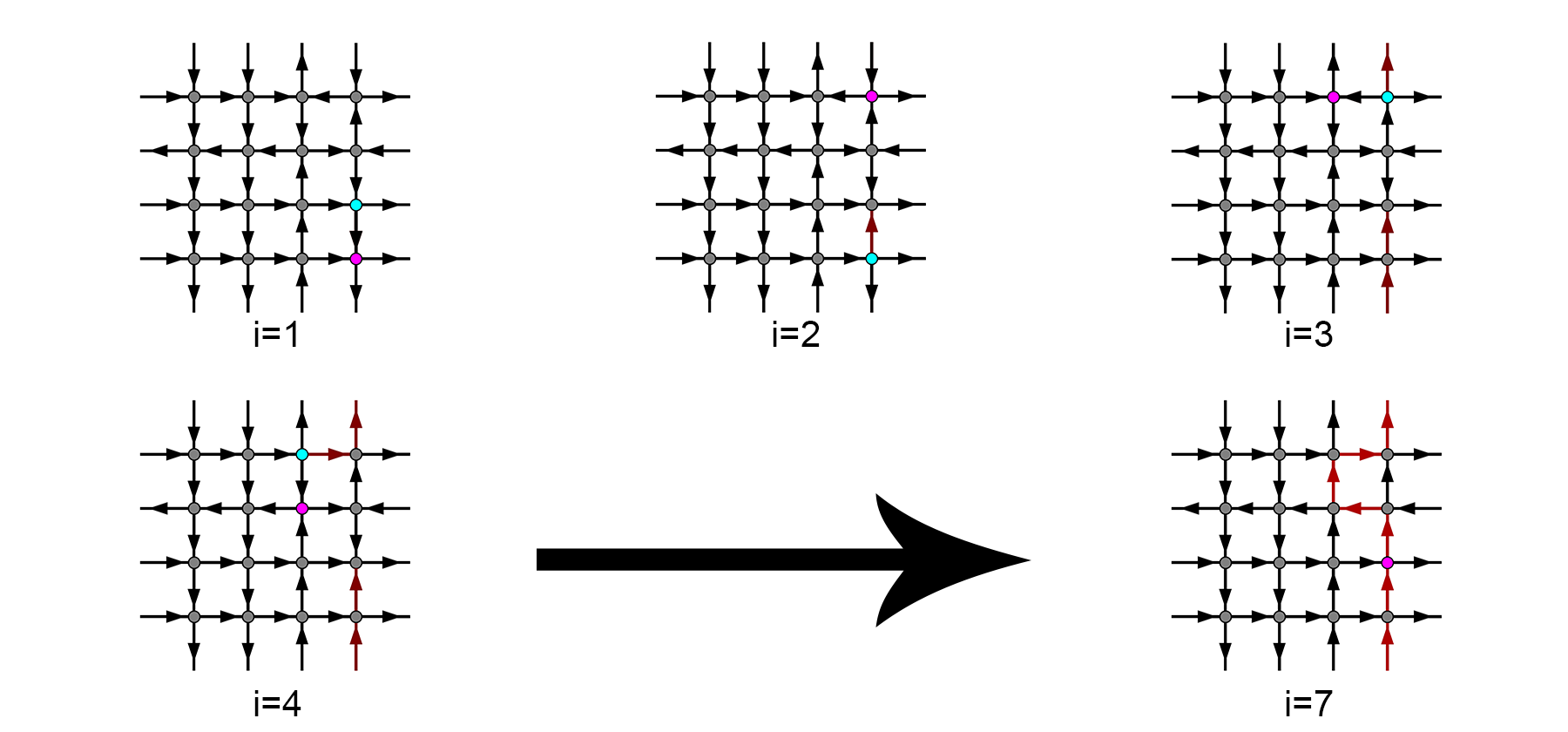}
	\caption{An example of a closed Long loop routine, gif available on \href{https://github.com/AKnightWing/ColdAsIce}{Github}}
	\label{fig:long_loop_report}
\end{figure*}

We will now consider using a Monte Carlo method to simulate an ice model. The first step in designing a Monte Carlo algorithm is to choose a set of moves that take us from one configuration to another, while making sure that both configurations satisfy the ice rules. This is what the Long loop algorithm does. The algorithm is as follows:
\newline \newline \textbf{Initialisation} 
\\ We start with a correct configuration of the model which satisfies the ice-rule constraint. \\
\newline \textbf{Main algorithm}
\begin{enumerate}
	
\item We choose a vertex at random from the lattice. Ice rules ensure that this vertex has two arrows going in and two arrows going out. This will be the starting point.
	
\item We choose at random one of the two outgoing arrows from this vertex and reverse it.
	
\item Now, the initial vertex has only one outgoing arrow instead of two and the new
vertex at the other other end of the reversed arrow has three. Clearly, the resulting configurations on both the vertices violate the second ice rule.
	
\item We then fix this new vertex by choosing one of the two outgoing arrows (ignoring
the one we just reversed) and reversing that. This fixes one vertex but creates a
new defect at the other end of the flipped arrow. We repeat this process to deal with
the next vertex.
	
\item  Thus, each time we repeat this process, it fixes one vertex but alters the other one. 

\end{enumerate}
\textbf{Termination}
\newline The process ends when we traverse a loop and return to the starting vertex. The configuration obtained at the end of these moves is one which satisfies the both the ice rules. Thus we have generated a new configuration of the system. 
\newline A working model of the Long loop algorithm is shown in Fig.(\ref{fig:long_loop_report})

The Long loop algorithm is ergodic and also satifies the condition for Detailed Balance\cite{Barkema_1998}.

\section{Symmetry Analysis}

Symmetry analysis is a crucial part of counting the number of states. If two seemingly different states have some symmetry transformation between them, then both those states correspond to the same unique state. In order to count the total number of unique states/configurations, we need to take a look at various symmetries generated by the long loop algorithm and eliminate any duplicates of a particular unique state that arise due to symmetry. 

There are three types of symmetries – \textit{rotation}, \textit{translation} and \textit{reflection}. We first look at the rotation and reflection symmetries. These correspond to the order of the Dihedral symmetry group of order $8$ (the symmetries of a square): $90^{\circ}$, $180^{\circ}$ and $270^{\circ}$ rotations about an axis that passes through the plane (z-axis), two $180^{\circ}$ flips about the horizontal axes (x and y-axes), two flips about the diagonal axes, and the identity transformation.

\begin{figure*}[h]
	\centering
	\includegraphics[scale=0.23]{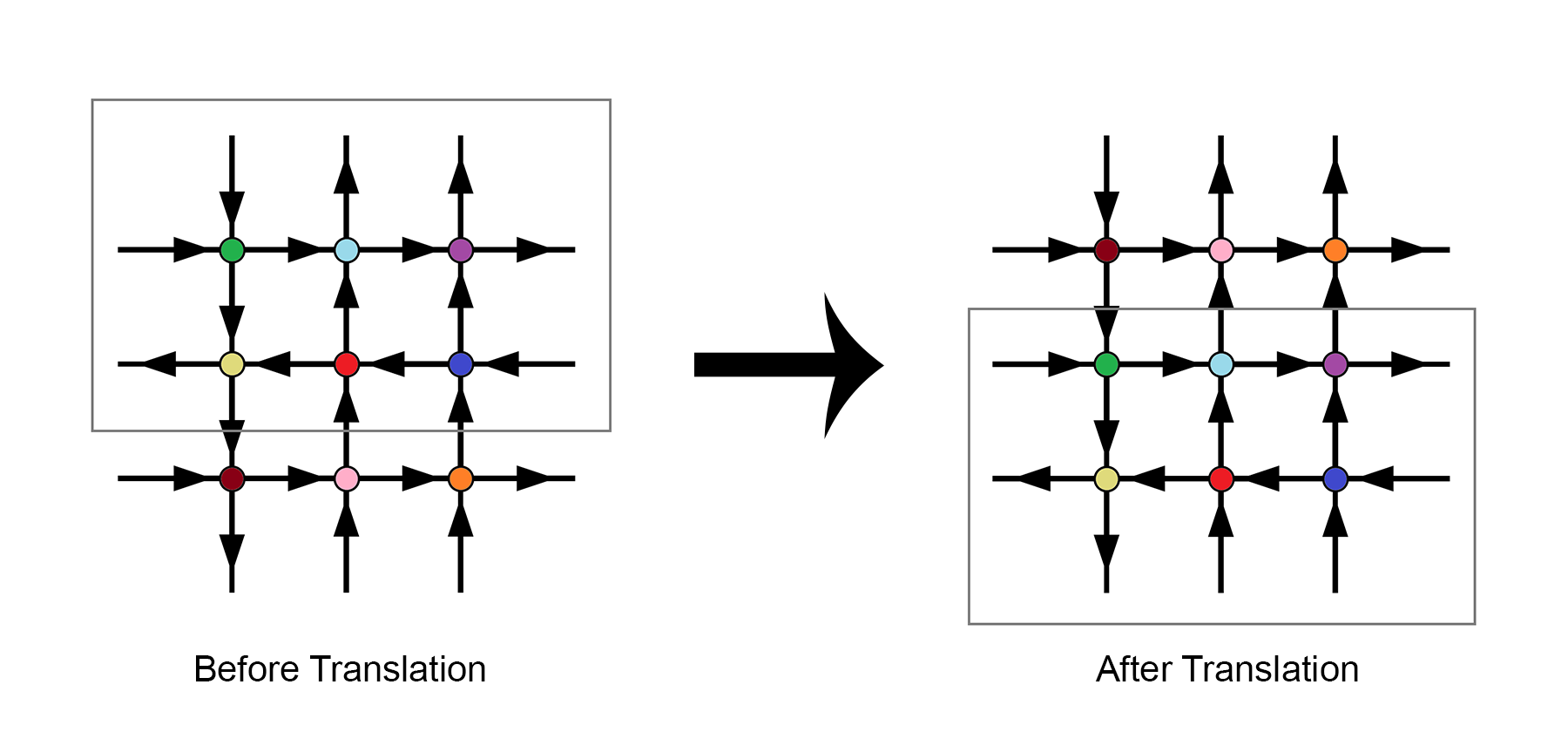}
	\caption{Translational symmetry of a configuration}
	\label{fig:translation}
\end{figure*}

The translational symmetries arise due to the periodic boundary conditions which we have imposed on the states. For example, if we translate a configuration by one step downwards, we would get a configuration that looks different but still corresponds to the same unique state, as shown in Fig.(\ref{fig:translation})

Periodic boundary conditions on all 4 sides of the ice map the 2D lattice onto a torus. Therefore, translation along any direction simply corresponds to a rotation of the torus along its primary axis.

All these symmetric states obey the ice rules and would be considered valid. This is fine if we simply wish to look at valid states but for the process of counting, it is important that we identify only one state and discard all of its symmetries.

\section{Counting States}
\begin{figure}[h]
	\centering
	\includegraphics[scale=0.45]{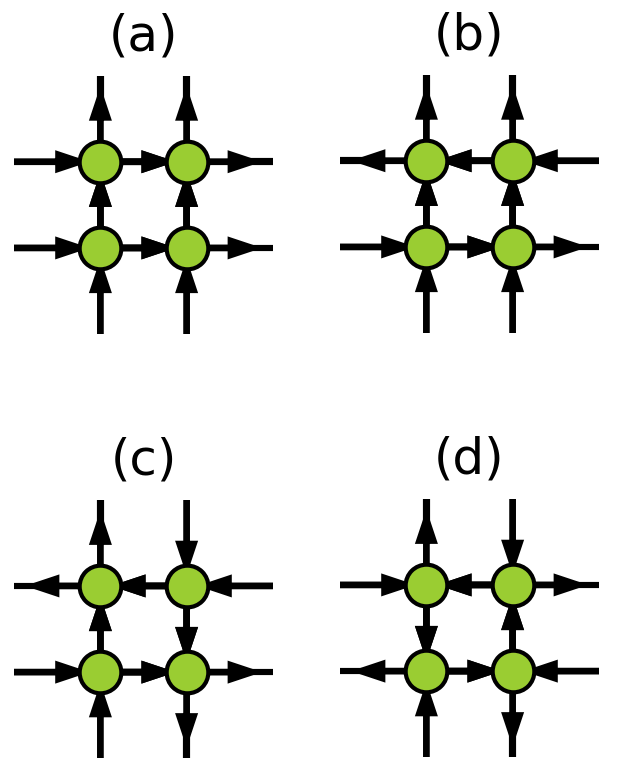}
	\caption{Unique Configurations of a 2x2 lattice}
	\label{fig:2x2}
\end{figure}

An immediate consequence of knowing the number of unique states of square ice is the knowledge of entropy of the system. 
\\To weed out symmetries and repetitions, we describe a notation that helps us identify every unique state.
We assign a string containing a sequence of $1's$ and $-1's$. This string can be generated by traversing through the array containing the state in a lexicographical order and appending the value to the string. For example, the identity for the state given in Fig.(\ref{fig:datarep}) is '$ 11-11111-1111-1111-1-11$'. This is guaranteed to generate a unique identity to every state except for symmetries. A dictionary is created that catalogues all these states. This will be used in the time stepping loop described below. 
The counting states algorithm is as follows:
\newline\newline\textbf{Initialization}
\newline Start with a configuration made entirely of Type 1 vertices. Use the Long loop algorithm on this configuration once. The output obtained is labelled as the initial state.
\newline\newline\textbf{Main algorithm}
\begin{enumerate}
\item Run the Long loop algorithm starting with the state generated at the previous step.
\item Generate the unique identification tag for the newly generated state.
\item Check through the dictionary containing all the unique states generated thus far by the long loop routine. If the current generated state is not present in the dictionary, it means that we have generated a new state and we append this to the dictionary. If the newly generated state is already present in this dictionary, we don’t do anything.
\item The process repeats by using the previously generated state as the new starting point for running the long loop routine again.	
\end{enumerate}
\textbf{Termination}
\newline The loop terminates when the approximate error reduces below a threshold.
\newline \textbf{Note} that approximate error is defined as
\begin{equation}
\textrm{Approximate Error} = \frac{\textrm{Good iterations}}{\textrm{Total iterations}} 
\end{equation}
where a Good iteration is when a Long loop routine arrives at a new state while a Bad iteration is when it does not. Total iterations is the sum of good and bad iterations.
\newline \newline At the end of this loop, we will have with us a dictionary that contains the identities of all unique states except for symmetries.
To remove symmetries we go through every state in the dictionary and find all possible rotational, translational and reflection symmetries for it described in Section 7.
We have written functions that generate these symmetries for any given input state. If any of these states are present in the dictionary, we delete it. We continue this process until we reach the last element of the dictionary.
At the end of this, we have a dictionary containing all the unique configurations of a lattice for a given number of atoms (for a 2x2 lattice, see Fig.(\ref{fig:2x2}) containing all 4 unique states). 
\section{Results: Non-energetic Ice}
\begin{figure}[h]
	\centering
	\includegraphics[scale=0.55]{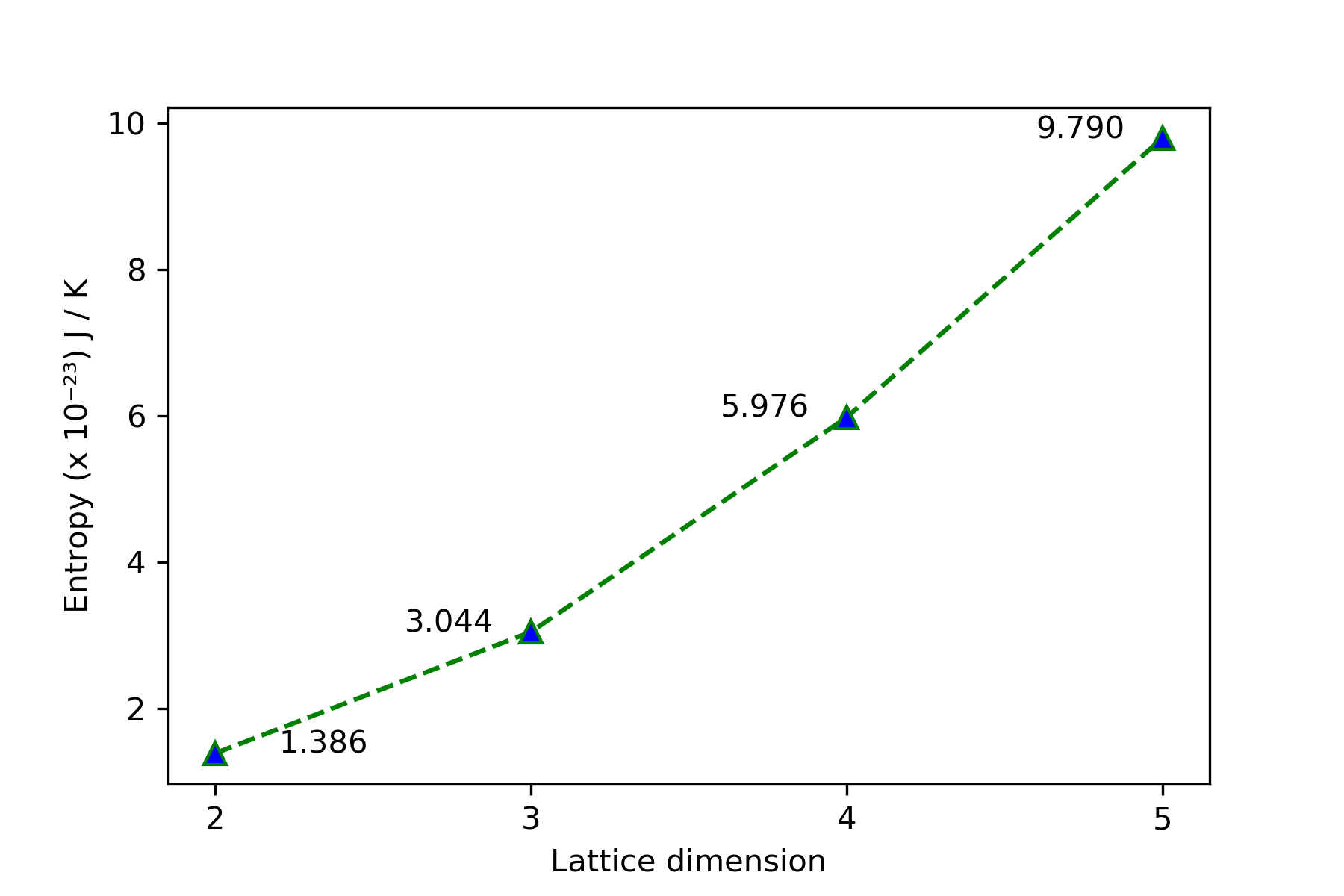}
	\caption{Entropy as a function of lattice dimension}
	\label{fig:entropy}
\end{figure}

\begin{figure}[h]
	\centering
	\includegraphics[scale=0.55]{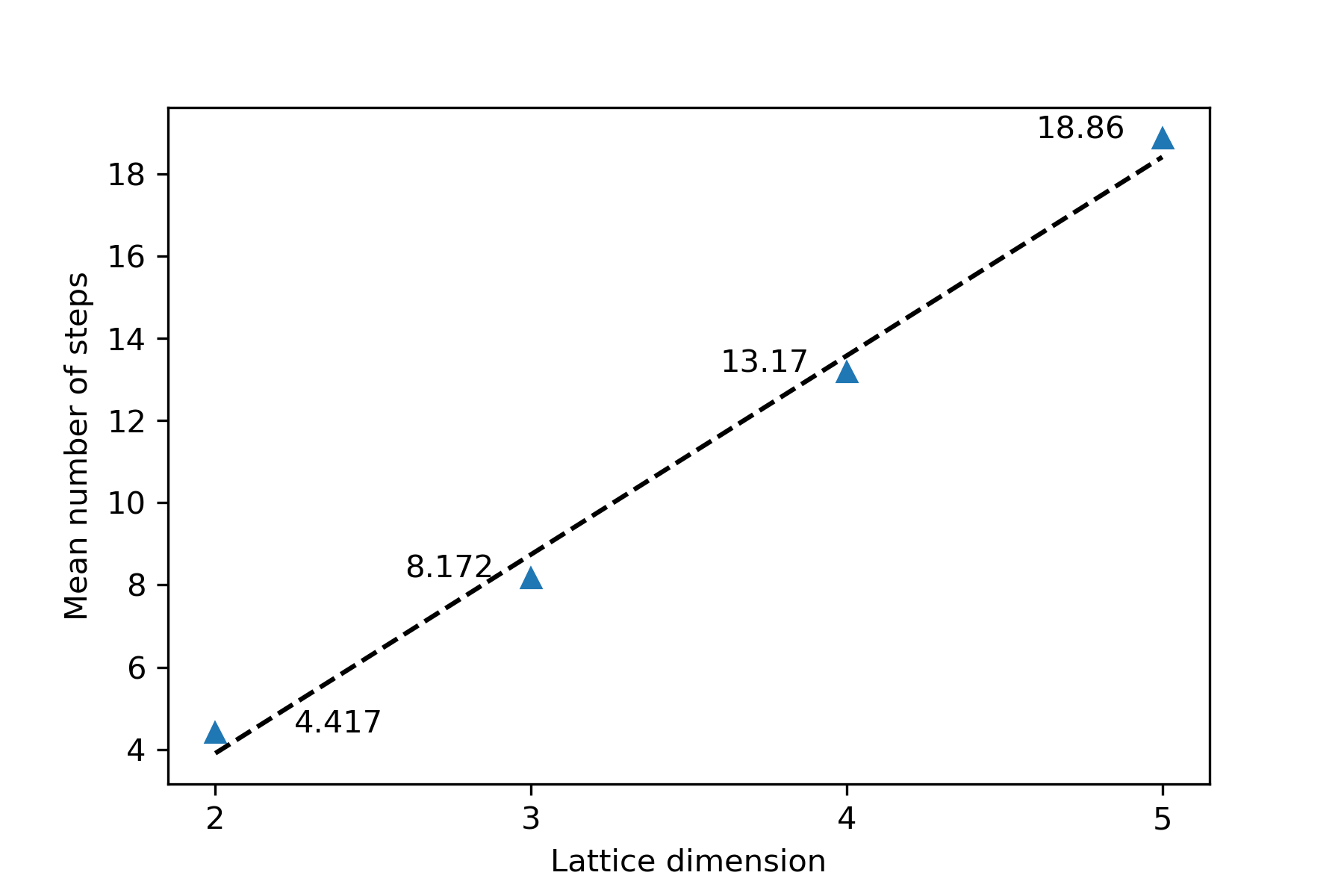}
	\caption{Mean number of Long loop steps v/s lattice dimension}
	\label{fig:mean_steps}
\end{figure}

\begin{figure}[h]
	\centering
	\includegraphics[scale=0.55]{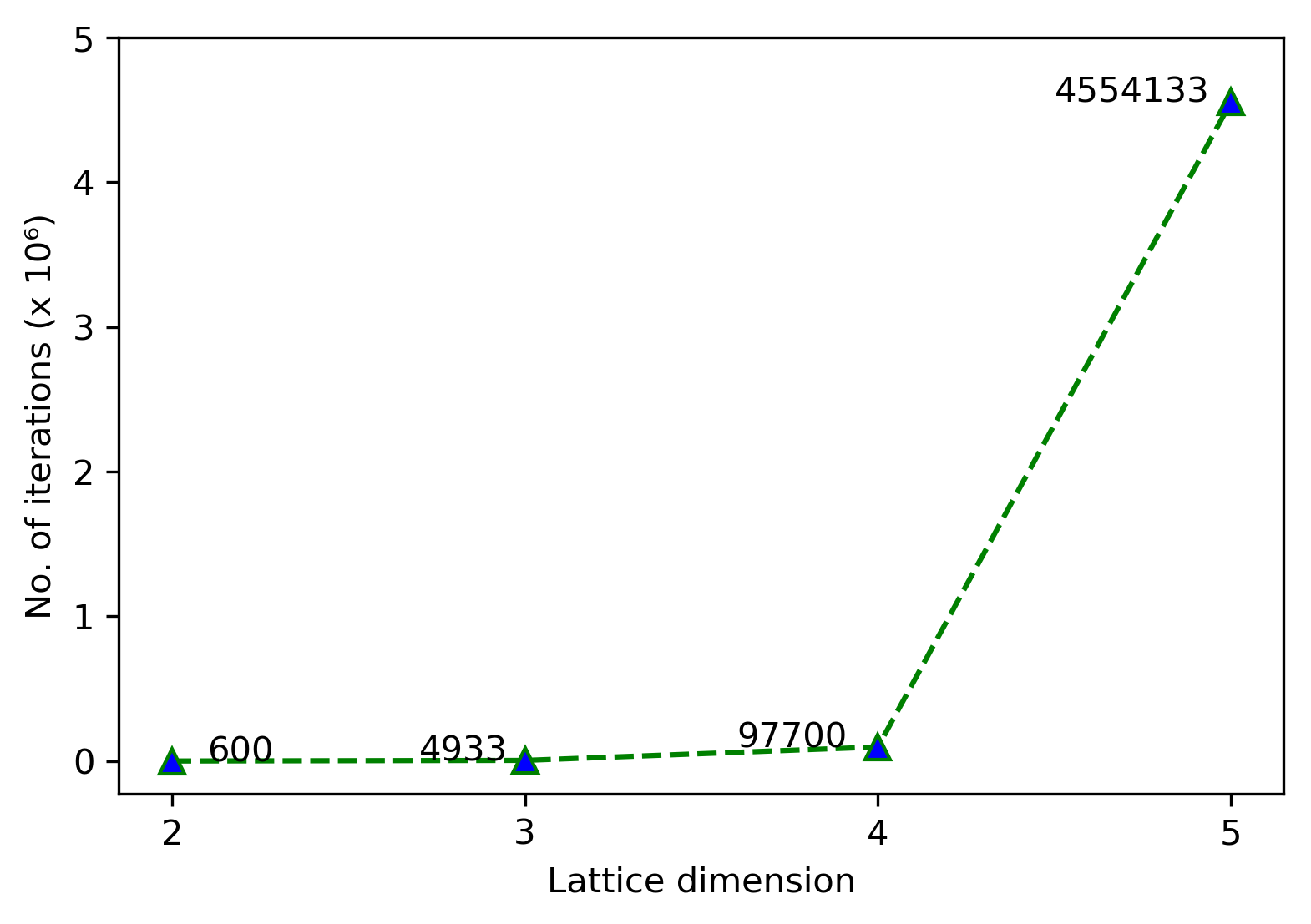}
	\caption{Number of Long loop iterations v/s Lattice dimension}
	\label{fig:num_iters}
\end{figure}

Entropy is calculated by counting the number of unique states (modulo symmetries) and is shown as a function of lattice dimension in Fig.(\ref{fig:entropy}).
Mean number of steps of a Long loop routine to arrive at a new state is also plotted as a function of lattice dimension in Fig.(\ref{fig:mean_steps}). \newline\textit{\textbf{Note}}: We find a best fit curve for Fig.(\ref{fig:mean_steps}) as
\begin{equation}
\textrm{mean steps} \propto L^{1.655}
\end{equation}
This matches Barkema et al's result for the same algorithm.\cite{Barkema_1998}
Total number of iterations required to converge to an approximate error of $3\%$ is also plotted as a function of lattice dimension in Fig.(\ref{fig:num_iters}). 

\section{Metropolis Algorithm}
The aim of our endeavour is to develop an algorithm to calculate thermodynamic quantities for ice at non-zero temperatures. For this, we need to simulate thermal fluctuations in a system. In reality, thermal fluctuations in a system are small compared to the entire energy of a system. To effectively sample a large number of states we consider setting the acceptance ratio:
\begin{equation}\label{boltzmann}
A(\mu \rightarrow \nu)=\left\{
\begin{array}{@{}ll@{}}
e^{-\beta(E_{\nu} - E_{\mu})}, & \text{if}\ \Delta E>0 \\
1, & \Delta E\leq 0
\end{array}\right.
\end{equation} 
\\The basis of Metropolis algorithm is as follows:
\newline\newline \textit{Given an initial state which satisfies the ice rules, we use the Long loop algorithm to generate a new state which also satisfies ice rules. 
\newline If this state has a lower energy than our previous state, we redefine this state as our initial state and run the process again.
\newline If this state has a higher energy than our previous state, we accept this state as our initial state with a Boltzmann probability given by Eq.(\ref{boltzmann}).}
\newline \textit{We see that the probability to accept a transition to a higher energy state decreases exponentially with lowering temperatures. }
\section{Equilibration}
Starting from an initial state, we need to run our simulation for a suitably long period of time until it has come to equilibrium at the temperature we are intersested in - this period is called \textbf{Equilibration time}. All thermodynamic calculations hereon are done after ensuring that the system has equilibrated.
\newline A mark of equilibration is that the order parameter of the system saturates to a fixed value. The order parameter of our system is \textbf{Polarization} or the average direction of arrows, defined as :
\begin{equation}\label{polarisation}
P = \frac{1}{\sqrt{2}N}\sum_{i}\hat{n}_i
\end{equation}
where $\hat{n}_i$ is a unit vector in the direction of the $i^{th}$ arrow.

\section{Energetic Ice }
An interesting extension of the ice models developed so far is to assign energies to different configurations. We mentioned in Section 3 that at absolute zero, all possible configurations have roughly the same energy. However, this energy changes as a function of temperature. The aim of this section will be to study the behaviour of energetic ice at non zero temperatures. To do this, we first need a convention to decide the energy of a given configuration.
\\We start with a model introduced by Franz Rys \cite{Rys_1963} called the \textbf{F model}. In this model, the symmetric vertices, given by Type 5 and Type 6 in Figure \ref{fig:6vertices}, are favoured by giving them a lower energy $-\epsilon$ while the rest of the vertices are given zero energy.

The Hamiltonian for the \textbf{F model} is then given by:
\begin{equation}\label{energy}
H = -\epsilon \sum_{i}\left[\delta_{v_i,5}+\delta_{v_i,6}\right]
\end{equation}
where $v_i$ is the type of vertex at site $i$, labelled according to our data representations scheme. 
\section{Fluctuations and Conjuagte Variables}
In classical thermodynamics; paramaters, constraints and fields interacting with the system each have conjugate variables which represent the response of the system to perturbations in the corresponding parameter. For example, the response of gas to a change in confining volume $V$ is a change in pressure $P$. Similarly, the magnetization $M$ of a magnet changes in response to the applied magnetic field $B$. Then $(P,V)$, $(M,B)$ are sets of conjugate variables. Classical thermodynamics also tells us that these conjugate variables are related to each other through derivatives of the Free energy $F$
\begin{equation}\label{conjugate}
\begin{split}
& P = -\frac{\partial F}{\partial V} \\ &  
M = \frac{\partial F}{\partial B}
\end{split}
\end{equation}
Note that $F = -k_B T \ln(Z)$.

We may extend the above argument for any given thermodynamic variable $X$, to generate an equation like Eq.(\ref{conjugate}) which links $X$ to its conjugate field $Y$. Derivatives of this general form are produced by a term in the Hamiltonian of the form $-XY$, where $Y$ is a "field" whose value we fix and $X$ is the conjugate variable to which it couples. 
\newline For example, the Hamiltonian of a magnet (consider the Ising model) has a term of the form $-MB$.
\newline The expectation value of $X$ can be written as:
\begin{equation}
\begin{split}
\langle X \rangle = & \frac{1}{Z} \sum_{\mu}X_{\mu}e^{-\beta E_{\mu}} \\ &
 = \frac{1}{\beta Z} \frac{\partial}{\partial Y}\sum_{\mu}e^{-\beta E_{\mu}}
\end{split}
\end{equation}
since $E_{\mu}$ contains a term of the form $-X_{\mu}Y$ which the $Y$ derivative acts on. 
\begin{equation}
\begin{split}
\langle X \rangle &= \frac{1}{\beta}\frac{\partial \ln Z}{\partial Y} \\ &
= - \frac{\partial F}{\partial Y} 
\end{split}
\end{equation}
\newline Now even if no field exists which couples to our variable $X$, we can simply make up a fictitious field $Y$ and add it to our Hamiltonian, perform the derivative to calculate the expectation of $X$, and then remove the term from the Hamiltonian again. 
\newline A second derivative of the partion function with respect to $Y$ produces:
\begin{equation}
\begin{split}
-\frac{1}{\beta}\frac{\partial^2 F }{\partial Y^2} &= \frac{1}{\beta}\frac{\partial\langle X \rangle }{\partial Y} \\ & 
= \langle X^2 \rangle - \langle X \rangle^2
\end{split}
\end{equation}
which we recognize as the mean square fluctuation in the variable $X$.
\newline The derivative $\partial \langle X \rangle/ \partial Y$, which measures the strength of the response of $X$ to changes in $Y$ is called the \textbf{susceptibility} of $X$ to $Y$.
\newline\newline The quantity of interest in our case is \textit{Polarization} $P$. The field that couples to it is an Electric Field $E$. Thus, we can define \textbf{Polarization Susceptibility} as:
\begin{equation}
\begin{split}
C_{P} & = \frac{\partial \langle P \rangle}{\partial E} \\ & 
= k_B \beta L^2 (\langle P^2 \rangle - \langle P \rangle^2)
\end{split}
\end{equation}
\subsection{Phase Transitions}
A phase of a system under consideration is characterised by the uniformity of certain physical properties throughout it. Two phases can be differentiated by looking at an \textbf{Order Parameter}, a quantity which takes two different values in different phases. In this case, the order parameter is \textbf{Polarization}, defined in Eq.(\ref{polarisation}). 
\\ Thermodynamic quantities show extreme behaviours at the critical point. For example, energy shows a discontinuous change at transition temperature, while generalised susceptibilities blow up.\cite{Callen}
\newline These behaviours cannot be reproduced accurately in numerical simulations. Instead, what we expect to see is a sharp change in energy near the critical point with a saturating behaviour on either side. 
\newline Similarly, generalised susceptibilities encounter their maxima (finite) at the critical point.
 
\section{Results: Energetic Ice}
We focus on three thermodynamic quantities: Energy, Specific Heat and Polarisation; each of which is calculated per unit number of atoms.\\
\textbf{We set $\mathbf{k_B = 1}$ to simplify calculations}.
Also, the value of $\epsilon$ is set to be 0.15. The metropolis algorithm has been iterated $10000$ times before calculating any thermodynamic quantity.

\subsection{Ensuring Equilibration}
\begin{figure}[h]
	\centering
	\includegraphics[scale=0.55]{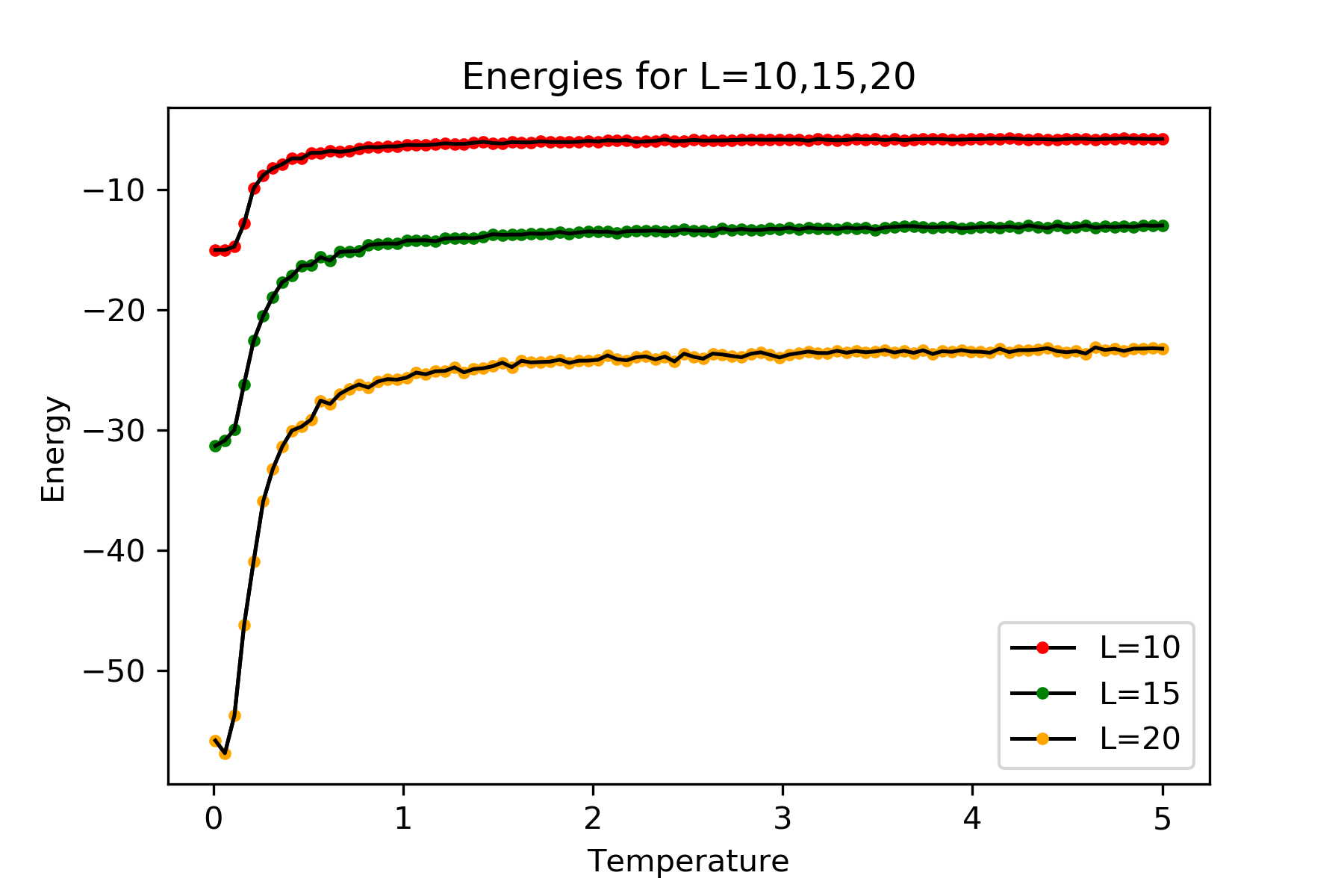}
	\caption{Energy v/s Temperature for different lattice sizes}
	\label{fig:energy}
\end{figure}

We find that 10,000 iterations are not enough to guarantee equilibration, especially at temperatures close to 0. This is because the Metropolis acceptance ratio for positive energy becomes exceedingly small at lower temperatures. This implies that for a large majority of Monte Carlo moves, the system only accepts flips to a lower energy state. Thus, the system tends to get stuck in a small region of the available phase space.
\newline An ideal scenario would be to run the metropolis algorithm for even longer; from our tests, we find that 750,000 iterations aree enough to guarantee equilibration. However, this requires a lot of computation power. 
\newline While 10,000 iterations aren’t always enough for equilibration, we theorize that there is a lower probability of the system requiring $>$10,000 iterations. From hereon, all the reported results have been obtained after taking a population mean over 15 identically generated samples for each temperature. This has helped in removing some of the pseudo-noise (due to the lower iteration size), and has helped us in characterizing the trends of the curve better. 
\newline \newline Now, Specific Heat of a system is related to fluctuations in energy as:
\begin{equation}
C = \frac{k_B \beta^2}{N}\left[\langle E^2\rangle - \langle E \rangle^2 \right]
\end{equation}
\newline Energy is calculated by averaging over 100 equilibrated states at each temperature. A Plot of energy as a function of temperature is shown in Fig.(\ref{fig:energy}) 
We observe a sharp change in energy in between two saturated values for all lattice sizes.
\subsection{L = 10}
\begin{figure}[h]
	\centering
	\includegraphics[scale=0.55]{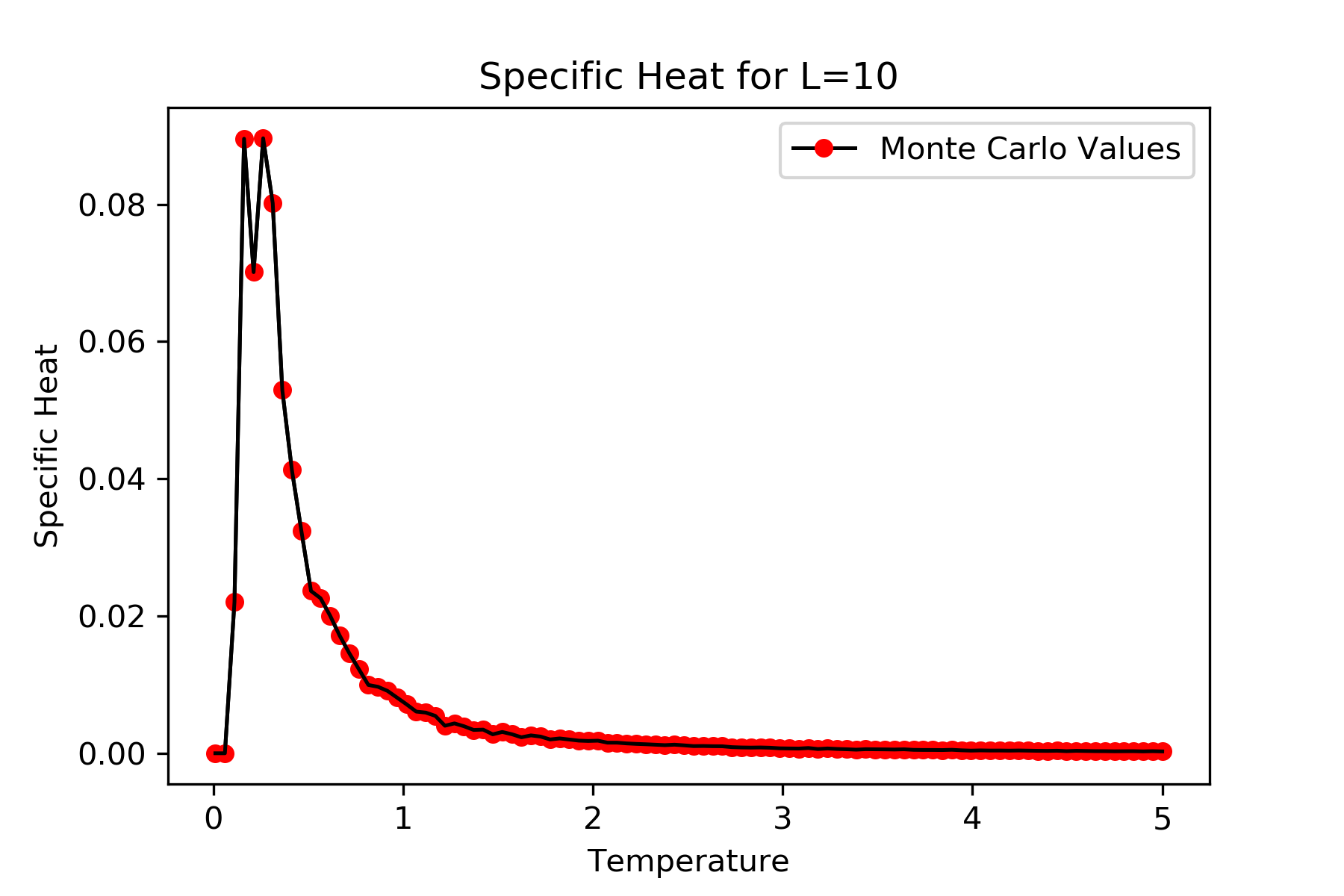}
	\caption{Specific Heat v/s Temperature for a $L=10$ Lattice}
	\label{fig:10SP}
\end{figure}
\begin{figure}[h]
	\centering
	\includegraphics[scale=0.55]{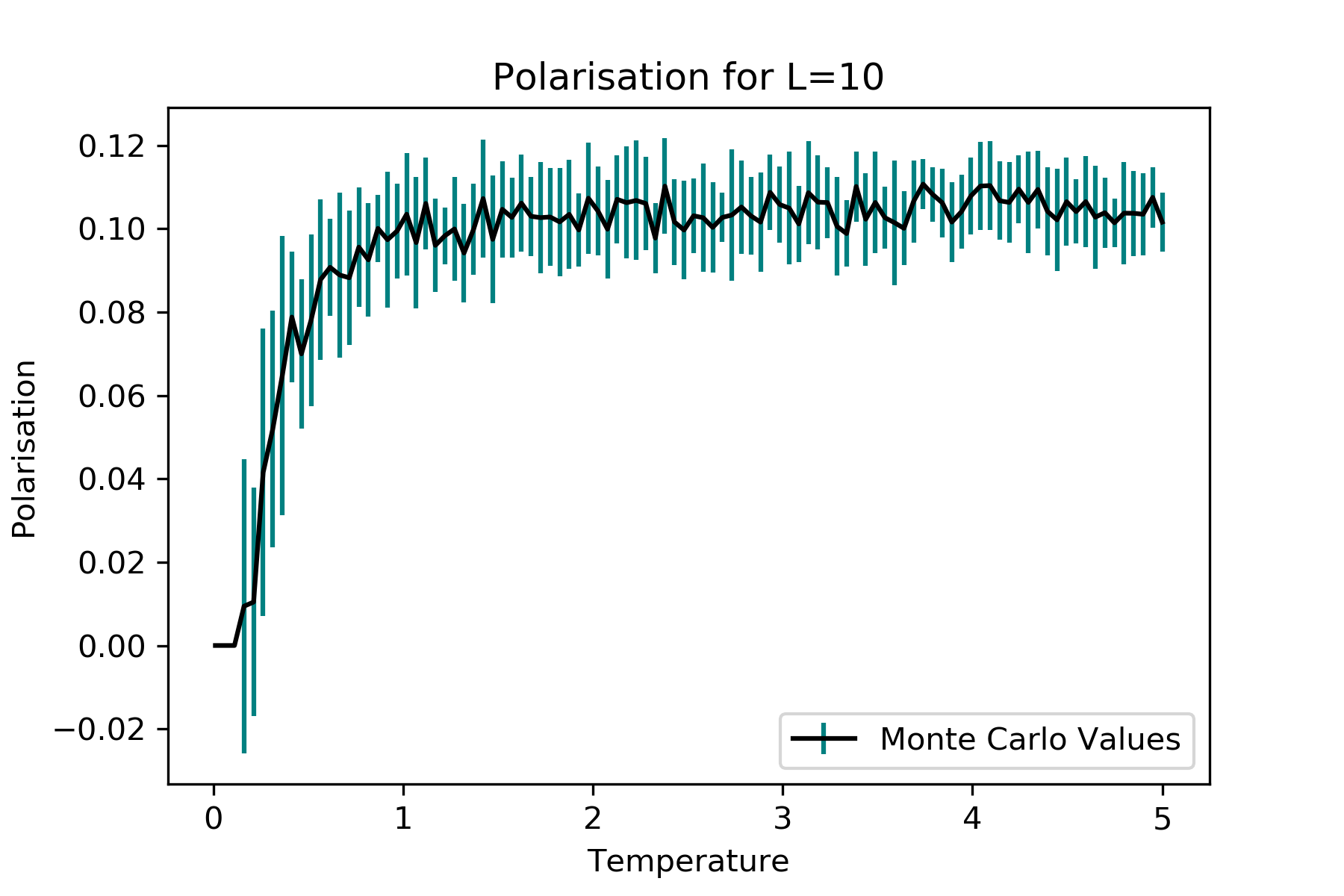}
	\caption{Polarization v/s Temperature for a $L=10$ Lattice}
	\label{fig:10P}
\end{figure}
\begin{figure}[h]
	\centering
	\includegraphics[scale=0.55]{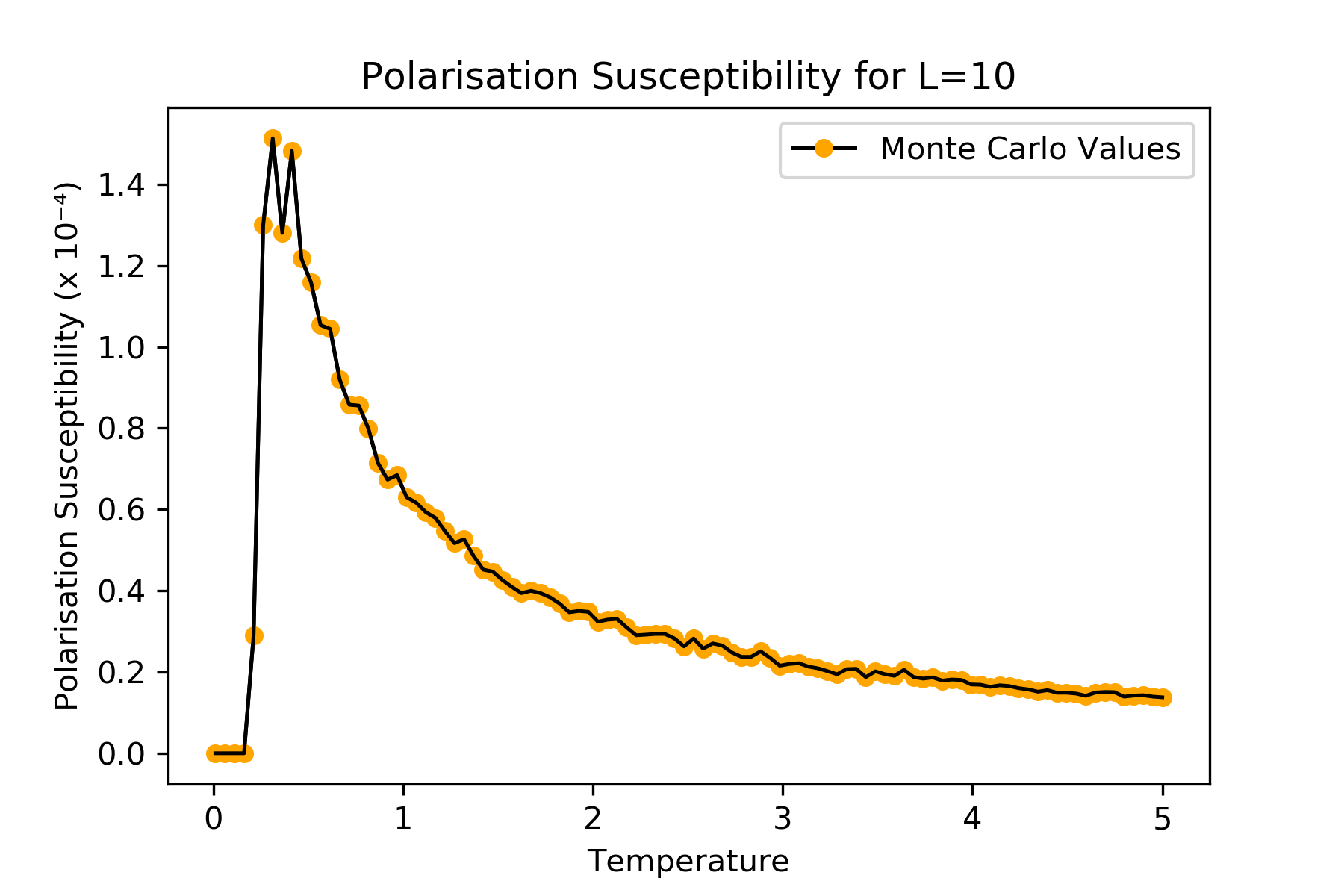}
	\caption{Polarization Susceptibility v/s Temperature for a $L=10$ Lattice}
	\label{fig:10PS}
\end{figure}
We also see that that Specific Heat for a $L = 10$ lattice attains its maxima in Fig.(\ref{fig:10SP}), which is also the temperature range in which energy goes through a change between two saturated values. 
\newline Polarization is plotted as a function of temperature with error bars in Fig.(\ref{fig:10P})
\newline Polarization Susceptibility also peaks and then diminishes for higher temperatures in Fig.(\ref{fig:10PS}). This peak occurs in the temperature range in which Polarization undergoes a sharp change in Fig.(\ref{fig:10P}).
\newline \textbf{This is evidence for a Phase transition in the system}.
\subsection{L = 15}
\begin{figure}[h]
	\centering
	\includegraphics[scale=0.55]{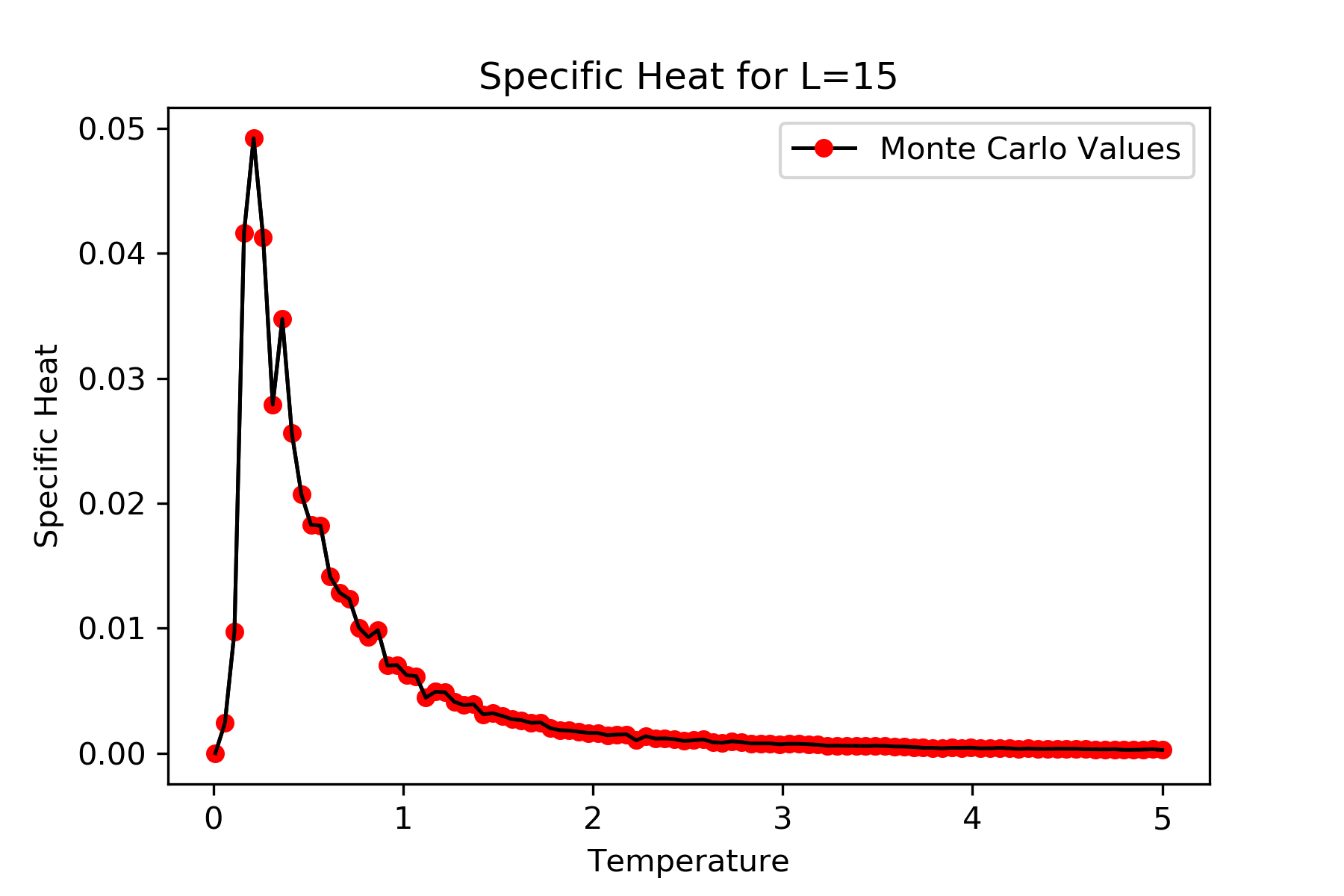}
	\caption{Specific Heat v/s Temperature for a $L=15$ Lattice}
	\label{fig:15SP}
\end{figure}

\begin{figure}[h]
	\centering
	\includegraphics[scale=0.55]{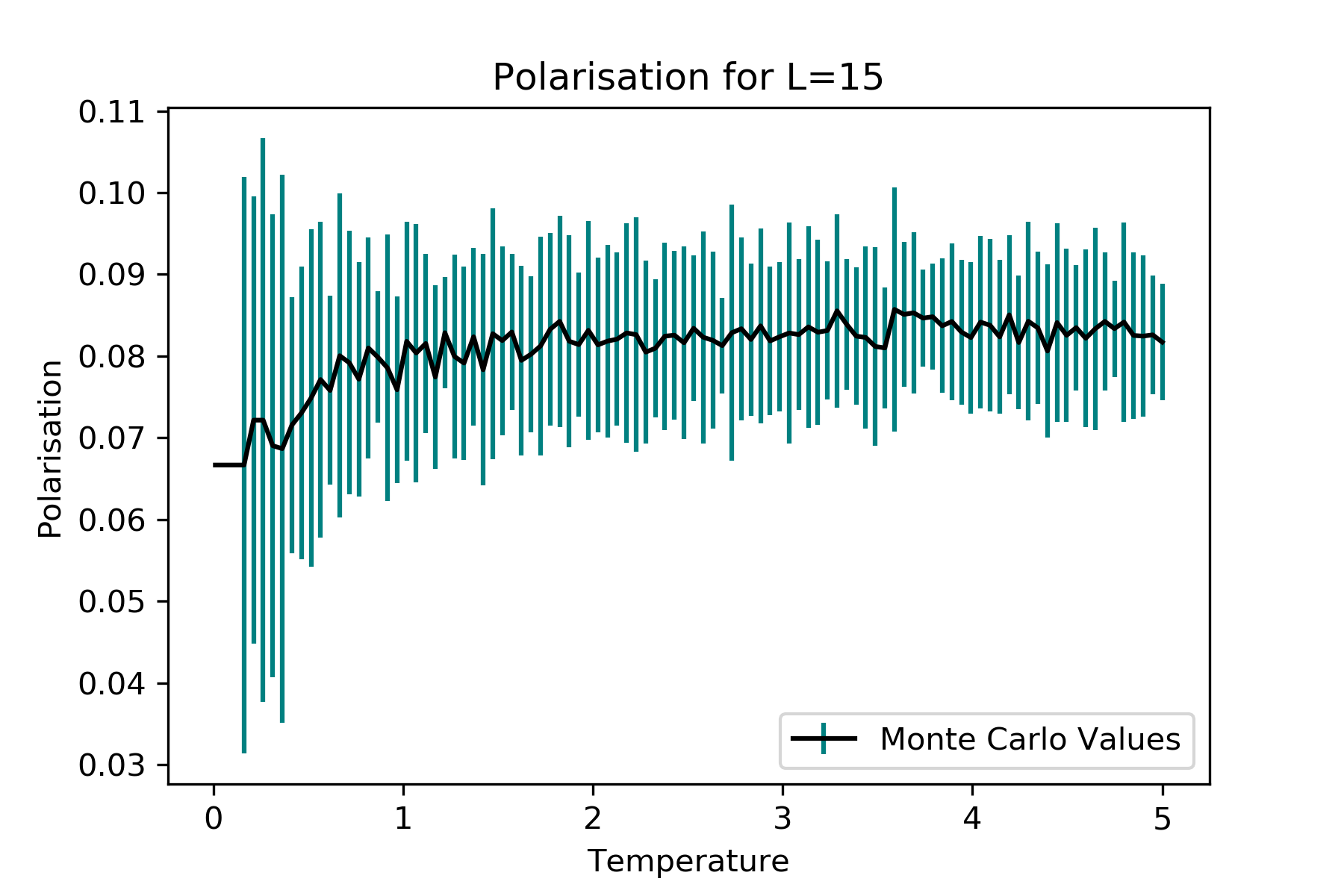}
	\caption{Polarisation v/s Temperature for a $L=15$ Lattice}
	\label{fig:15P}
\end{figure}
\begin{figure}[h]
	\centering
	\includegraphics[scale=0.55]{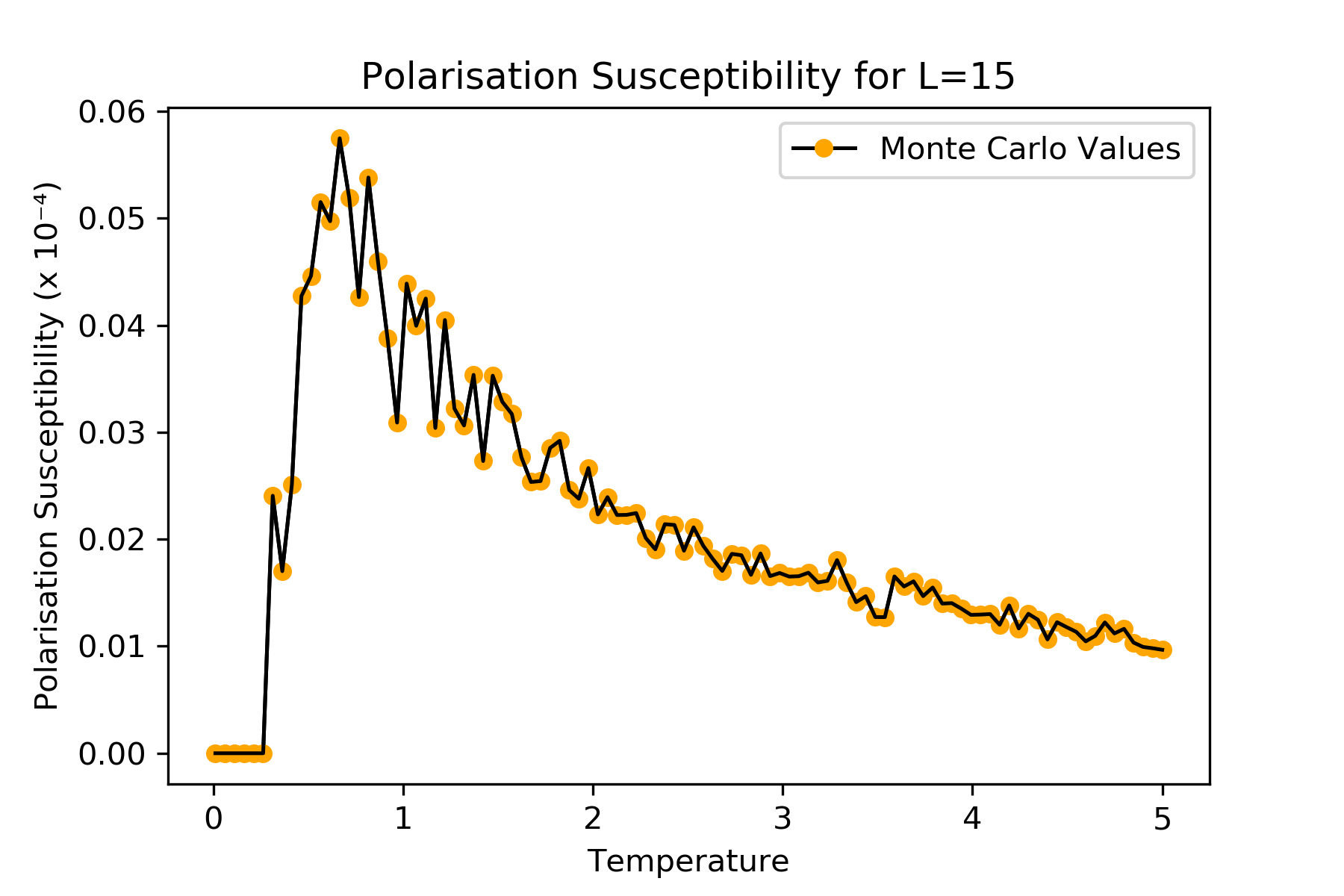}
	\caption{Polarisation Susceptibility v/s Temperature for a $L=15$ Lattice}
	\label{fig:15PS}
\end{figure}
For $L=15$, we observe behaviour similar to the $L=10$ case in Fig.(\ref{fig:15SP}), Fig.(\ref{fig:15P}) and Fig.(\ref{fig:15PS}). 

\subsection{L = 20}
\begin{figure}[h]
	\centering
	\includegraphics[scale=0.55]{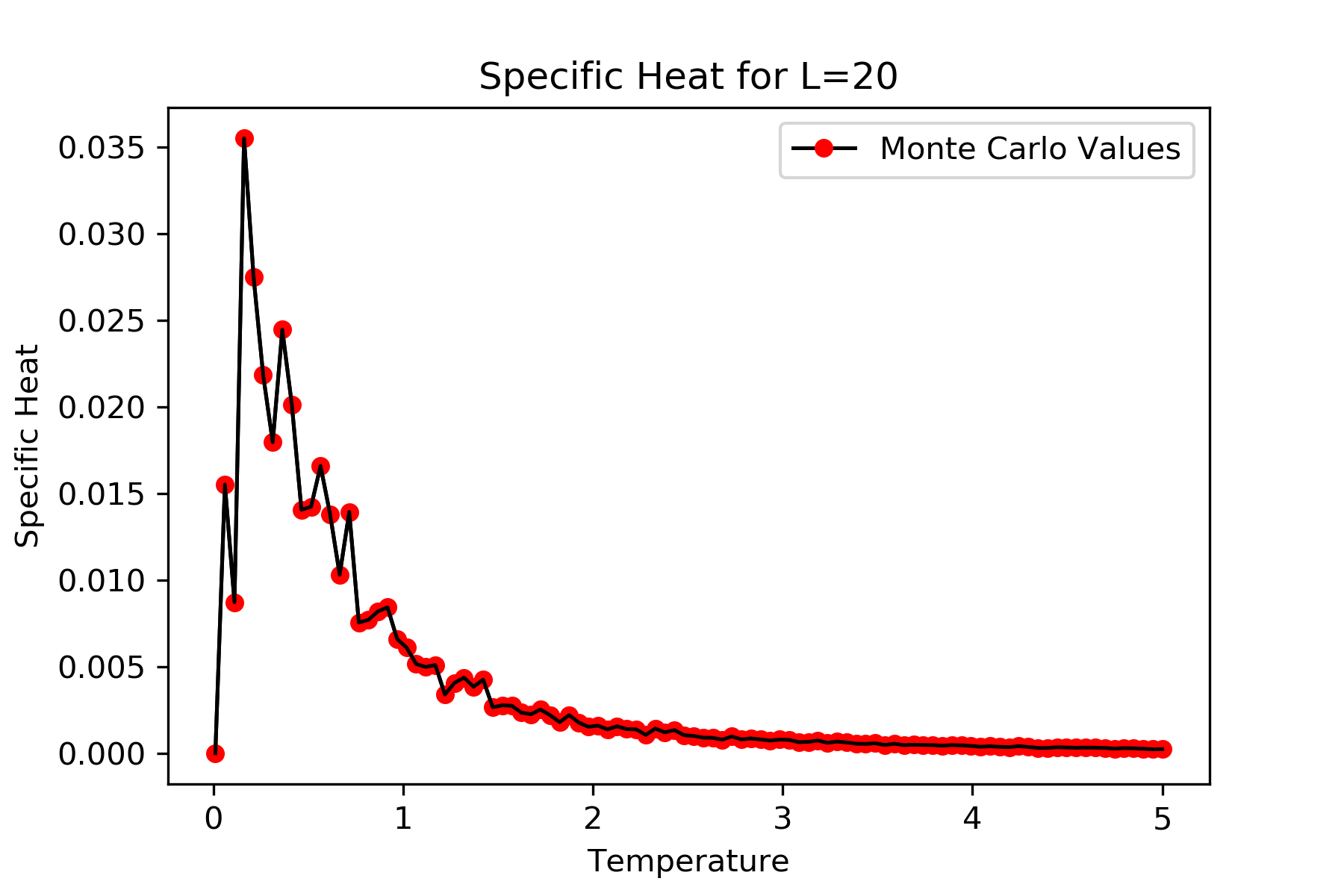}
	\caption{Specific Heat v/s Temperature for a $L=20$ Lattice}
	\label{fig:20SP}
\end{figure}

\begin{figure}[h]
	\centering
	\includegraphics[scale=0.55]{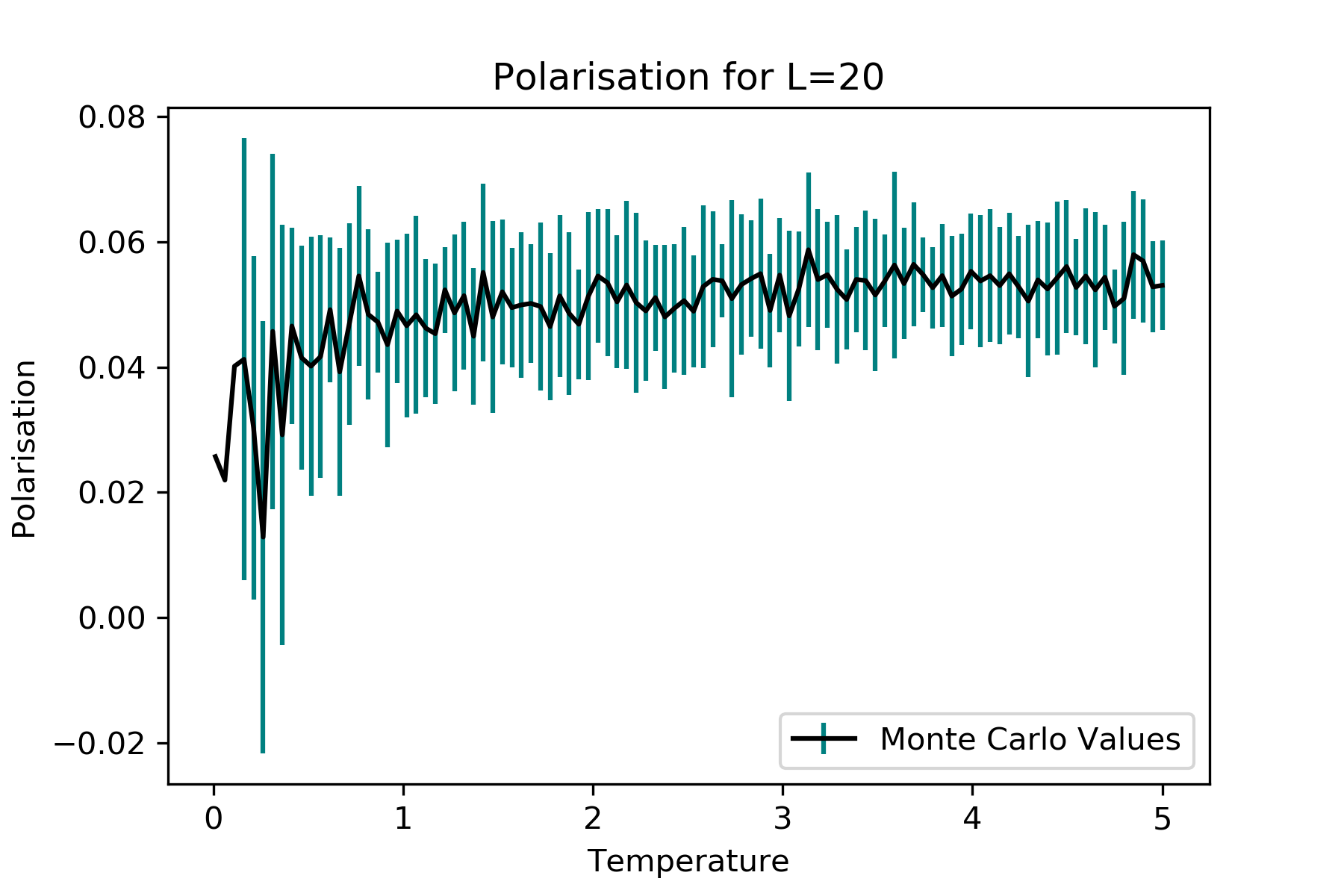}
	\caption{Polarisation v/s Temperature for a $L=20$ Lattice}
	\label{fig:20P}
\end{figure}
\begin{figure}[h]
	\centering
	\includegraphics[scale=0.55]{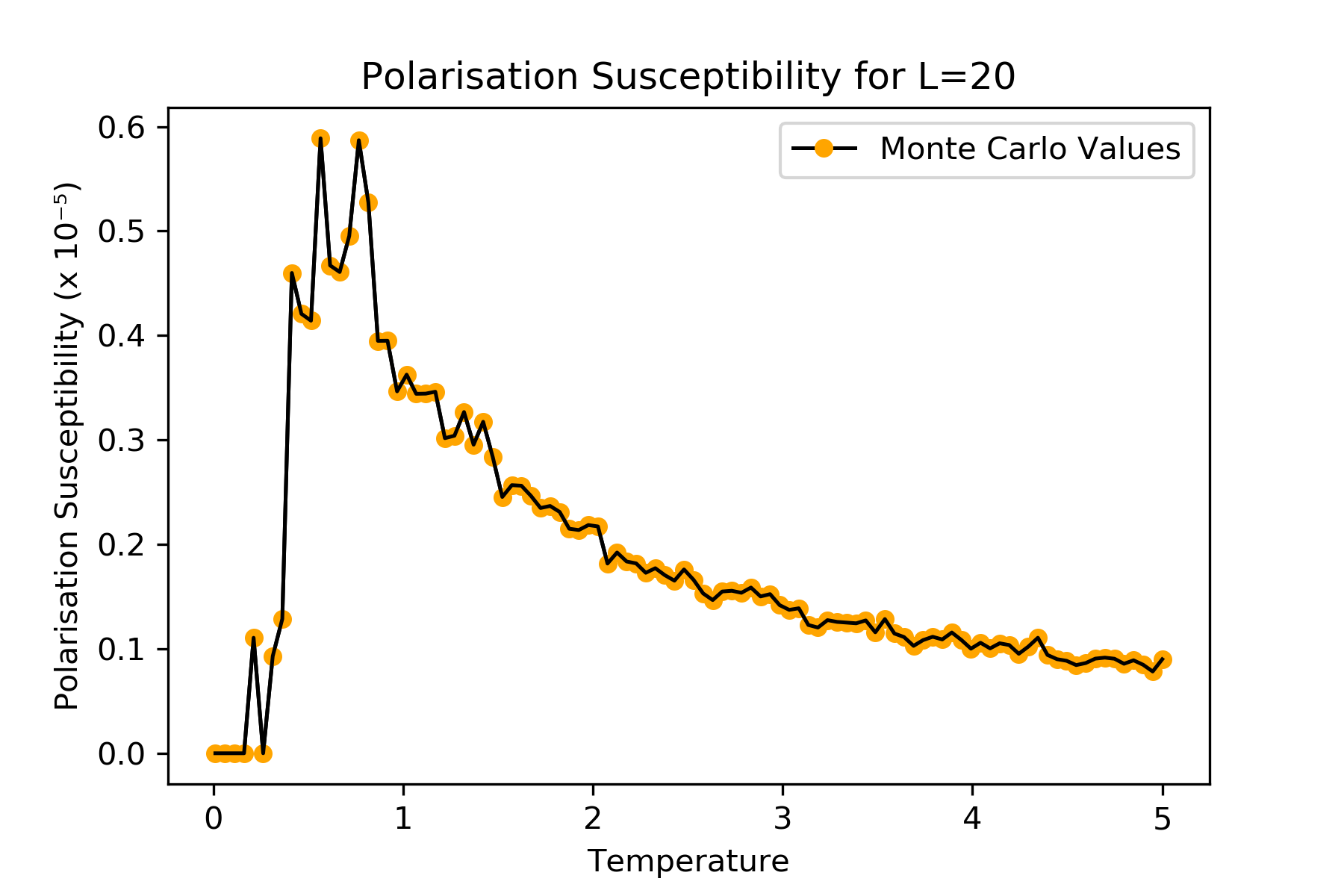}
	\caption{Polarisation Susceptibility v/s Temperature for a $L=20$ Lattice}
	\label{fig:20PS}
\end{figure}

The trend continues for higher lattice dimensions, with peaks in Polarization susceptibility and Specific Heat falling in the same temperature range as discontinuous changes in Polarization and Energy respectively. 
\newline \textbf{The occurence of this behaviour for all 3 lattice sizes strongly hints at the presence of a phase transition in the system. This forms the central result of our text}.

\subsection{Equilibrium Configurations}
The \textbf{F model} favours Type 5 and Type 6 vertices by giving them a lower energy than the rest. These 2 vertices can cover the 2-D plane by themselves. As the ground state of the system is the state of least energy at any temperature, the least energy configuration obtained at absolute zero exclusively consists of Type 5 and Type 6 vertices arranged in a checkerboard pattern.
\newline The accuracy of the Metropolis algorithm can be seen from the acceptance ratios defined in Eq.(\ref{boltzmann}). The probability of acceptance of a state at higher energy becomes 0 at absolute zero. Thus no iteration of the algorithm results in an increased energy of the configuration. 
\newline Note that there are two possible ground states, depending on which vertex(5 or 6) falls on an even site and which falls on the odd. At higher temperatures, both of these vertices are equally likely to fall into even and odd sites.
\newline Thus, \textit{on lowering the temperature the system undergoes a phase transition into one of the states. (In which Type 5 occupies odd sites and Type 6 occupies even sites or vice versa)}.
\newline With increasing temperatures, the system moves to states of higher energies, and hence we expect to see a higher concentration of Type(1-4) vertices. 

\begin{figure*}[h]
	\centering
	\includegraphics[scale=0.33]{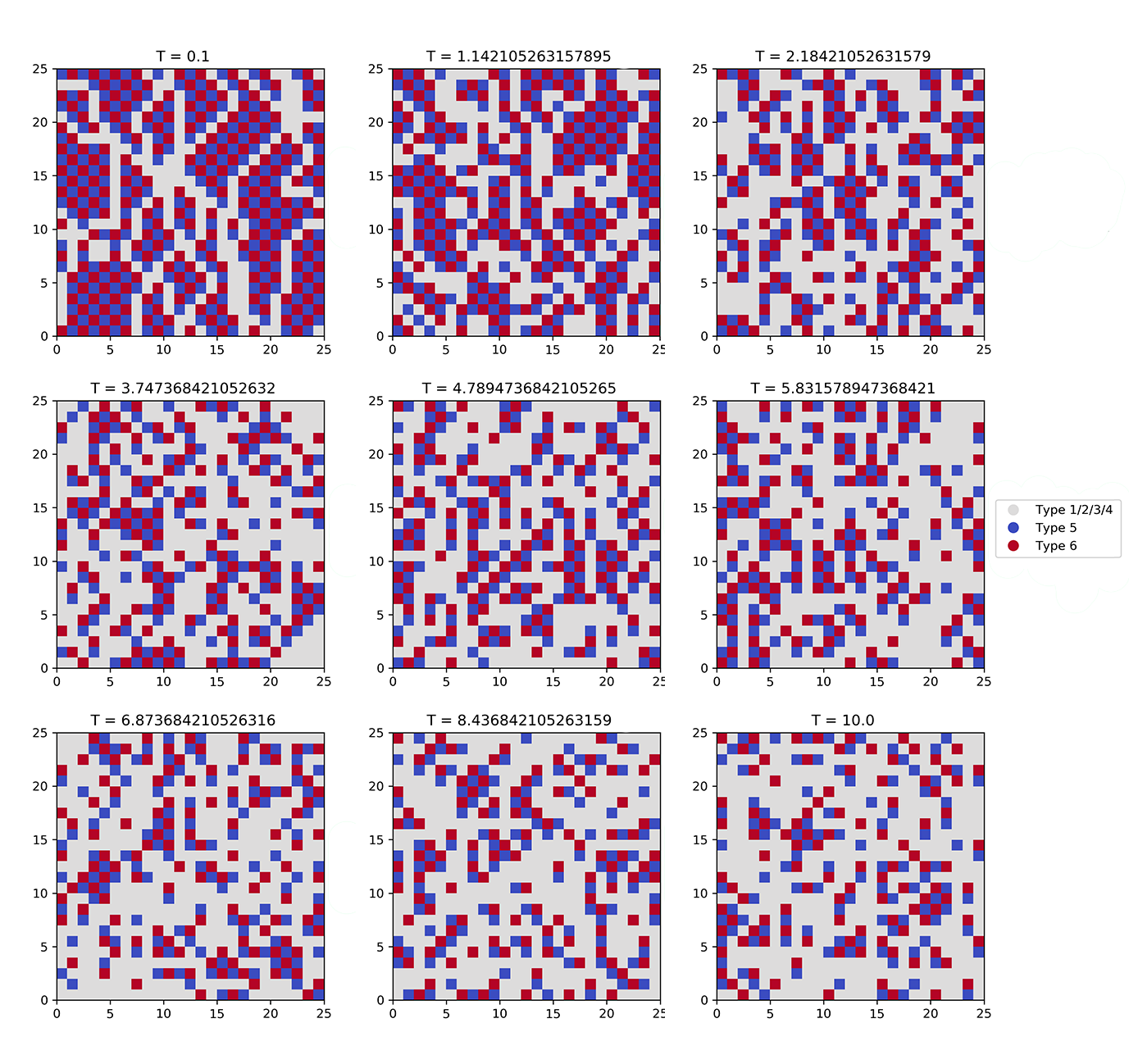}
	\caption{Equilibrated configurations at different temperatures, GIF available on \href{https://github.com/AKnightWing/ColdAsIce}{Github}}
	\label{fig:temp_snapshots}
\end{figure*}

Fig.(\ref{fig:temp_snapshots}) shows snapshots of configurations at various temperatures ranging from $T=0.1$ to $T=10.0$. These have been obtained by running the Metropolis algorithm till equilibration is achieved. We clearly see a decreasing concentration of Type (5-6) vertices and a corresponding increase in concentration of Type(1-4) vertices.
\section{Scope for future work}
Our immediate aim is to calculate finite temperature entropy for energetic ice by integrating the specific heat data generated. We also plan to study a 3D model of energetic ice. By implementing our algorithms for a pyrochlore lattice, we also plan to reproduce the work done by Ramirez et al.\cite{spin_ice} on spin ice systems.
\section{Aknowledgements}
We thank Dr.~Sunil Pratap Singh for suggestions on energetic ice models. We are also grateful to Dr.~Auditya Sharma and Mr.~Vedula Bharadwaj for their insights on Phase Transitions. 
We are thankful to \href{https://www.kaggle.com/}{Kaggle} for providing computational resources.

\bibliographystyle{unsrt}
\bibliography{references}

\end{document}